\documentclass{desyproc}

\begin{document}
\title{Resonance Searches at HERA}
 
\author{{\slshape Uri Karshon$^1$}\\[1ex]
$^1$Weizmann Institute of Science, Rehovot, Israel\\
On behalf of the H1 and ZEUS
Collaborations }  
 
\contribID{79}
 
\confID{1407}  
\desyproc{DESY-PROC-2009-03}
\acronym{PHOTON09} 
\doi  
 
\maketitle
 
\begin{abstract}
   
Inclusive production of $K^0_S K^0_S$ in ep collisions         was studied with the
ZEUS detector. Significant production of     $J^{PC}=2^{++}$ tensor mesons                                       
and of the $0^{++}$ glueball candidate $f_0(1710)$
was seen. Masses and widths were compared with
previous experiments.
                                                                         The H1
Collaboration saw a charm pentaquark candidate in the $D^* p$ spectrum at
$3.1$~GeV, which was not confirmed by a ZEUS higher statistics search.                         
        With the full HERA statistics, H1 did not see a signal in this region.             
Masses, widths and helicity parameters of
                  excited charm and charm-strange mesons were measured by ZEUS.
                                          Rates of $c$ quarks hadronising into these  
              mesons were determined and a search for a radially excited charm meson
was performed.                                                                      
\end{abstract}
 
\section{Introduction}
 
The HERA $e p$ collider operated with electrons or positrons at 27.6~GeV and protons
at 820 or 920~GeV. 
         Each of the two general purpose experiments H1 and ZEUS collected 
during 1995 - 2000 (``HERA~I") 
 $\approx 120~pb^{-1}$ and during 2003 - 2007 (``HERA~II") $\approx 370~pb^{-1}$.
 Two kinematic regions have been explored: Deep inelastic scattering (DIS)
with photon virtuality  $Q^2 > 1$~GeV$^2$, where the scattered electron is visible in 
the main detector and photoproduction (PHP) with $< Q^2 > \approx 3\cdot 10^{-4}$~GeV$^2$,
where the virtual photon radiated from the incoming electron is quasi-real.
The sample is dominated by PHP events.
 



\section{Glueball search in the $K^0_S K^0_S$ system}
 

Glueballs are predicted by QCD. The lightest glueball is expected to have 
$J^{PC}=0^{++}$ and a mass in the range 1550-1750~MeV~\cite{PDG} and    can
mix with $q\bar q$ scalar meson nonet I=0 states of similar mass. There are four
such established states: $f_0 (980)$, $f_0 (1370)$, $f_0 (1500)$ and $f_0 (1710)$,
but only two can fit into the nonet. The  $f_0 (1710)$ state is considered as a possible
glueball candidate. The $K^0_S K^0_S$ system can couple to $J^{PC}=0^{++}$ and $2^{++}$.
Therefore, it is a good place to search for the lowest lying $0^{++}$ glueball.
 
\subsection{Previous results}

The  $e^+ e^-$ experiments TASSO and L3 studied the exclusive reaction $\gamma\gamma\to K^0_S K^0_S$.
L3~\cite{L3} saw 3 peaks and attributed them to $f_2(1270)/a_2(1320)$, $f^{'}_2(1525)$ and $f_0(1710)$.
A maximum likelihood fit with 3 Breit-Wigner (BW) functions plus background yielded     $f^{'}_2(1525)$
mass and width values consistent with the Particle Data Group (PDG)~\cite{PDG} and a 4 standard deviation
               (s.d.) signal for $f_0(1710)$ with mass and width
values above     PDG. The TASSO~\cite{TASSO} $K^0_S K^0_S$ spectra had no $f_2(1270)/a_2(1320)$ signal and
a sizable $f^{'}_2(1525)$ enhancement. The result was interprated by interference effects between
the 3 $J^P =2^+$ resonances $f_2(1270)$, $a_2(1320)$ and $f^{'}_2(1525)$ and the spectra was fitted
as a sum of 3 coherent BW functions. Based on SU(3) symmetry arguments~\cite{faiman}, the sign of the
$a_2(1320)$ term for $K^0_S K^0_S$ is negative and the coefficients of the 
$f_2(1270)$, $a_2(1320)$ and $f^{'}_2(1525)$ BW amplitudes are       +5, -3 and +2, respectively.
 
 
\subsection{This analysis}
 
%
 \begin{wrapfigure}{r}{0.5\columnwidth}
\centerline{\includegraphics[width=0.40\textwidth]{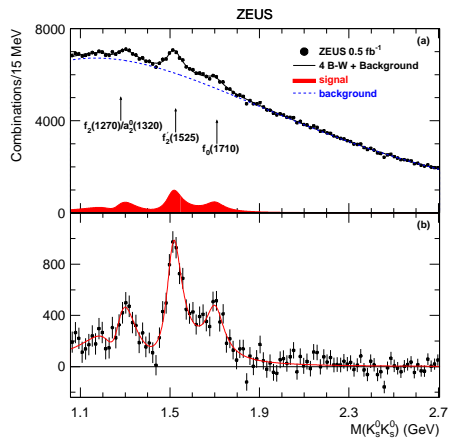}}
\caption{(a)The $K^0_S K^0_S$  distribution (dots). 
Solid line is the coherent fit (see text);        
background function is given by the dashed line.                               
(b)Background-subtracted $K^0_S K^0_S$                                    
distribution (dots); solid line is the fit result.}
                                               
                    \label{Fig:ksksco}
                                                
\end{wrapfigure}

                                              The reaction            $e^{\pm}p\to K^0_S K^0_S + X$ 
was studied~\cite{ksks}                                                    
                          with the full HERA luminosity of 0.5 fb$^{-1}$.      Both PHP 
and DIS events were included.
No explicit trigger requirement was applied for selecting the above reaction.                  

$K^0_S$ mesons were identified via their decay mode $K^0_S\to\pi^+\pi^-$.
 A clean $K^0_S$ signal was seen     
                                          for events with
$\ge 2 K^0_S$ candidates.                                                                   
                          The number of $K^0_S K^0_S$ pairs found in the $K^0_S$  mass range 
$481 < M(\pi^+\pi^-) < 515$~MeV is $\approx 672,000$.  

Figure \ref{Fig:ksksco} shows 
the $K^0_S K^0_S$ mass distribution reconstructed by combining two $K^0_S$ candidates selected
in the above mass window. Three peaks are seen around 1.3, 1.5 and 1.7 GeV.  
                   No state heavier than
1.7 GeV was observed. The invariant-mass spectrum, $m$, was fitted as a sum of relativistic         
Breit-Wigner (RBW) resonances              
and a smoothly varying background $U(m)=m^A exp(-Bm)$, where A and B are free parameters.
 
 
Two types of fit, as performed for the reaction $\gamma\gamma\to K^0_S K^0_S$ by L3~\cite{L3} and
TASSO~\cite{TASSO}, respectively, were tried. The first fit (not shown)   
        is an incoherent sum of three modified
RBW resonances, $R$, of the form
$F(m)=C_R(\frac{M_R\Gamma_R}{(M_R^2 - m^2)^2 + M_R^2\Gamma_R^2})$, representing the peaks
$f_2(1270)/a_2(1320)$, $f^{'}_2(1525)$ and $f_0(1710)$. Here $C_R$ is the resonance amplitude and
$M_R$ and $\Gamma_R$ are the
resonance mass and width, respectively.
The goodness of this fit is reasonable ($\chi^2 /ndf = 96/95$); however, the dip between the 
$f_2(1270)/a_2(1320)$ and  $f^{'}_2(1525)$ is not well reproduced.


Figure \ref{Fig:ksksco} shows a coherent fit motivated by SU(3) predictions\cite{faiman}. 
Each resonance amplitude, $R$, is described by the RBW form~\cite{TASSO}
$BW(R)=\frac{M_R\sqrt{\Gamma_R}}{M_R^2 -m^2 -iM_R\Gamma_R}$.
The decays of the tensor ($J^P =2^+$) mesons $f_2(1270)$, $a_2^0(1320)$ and $f^{'}_2(1525)$
into the two pseudoscalar ($J^P =0^+$) mesons $K^0\bar K^0$ are related by SU(3) symmetry with
a specific interference pattern. The intensity is the modulus-squared of the sum of these 3
amplitudes plus the incoherent addition of $f_0(1710)$ and a non-resonant background.

 
 
Assuming SU(3) symmetry and a direct coupling of the $2^+$ states to the exchanged photon, 
the fitted function to the $m(K^0_S K^0_S)$ spectra             
                                   is given by                 
$F(m) = a[5\cdot BW(f_2(1270)) -                   
          3\cdot BW(a_2(1320)) +   
          2\cdot BW(f^{'}_2(1525))]^2  
      + b[BW(f_0(1710))]^2  +      
        c\cdot U(m)$, where a,b,c as well as
    the resonance masses and widths were free parameters in the fit.
The background-subtracted mass spectrum is shown in Fig.\ref{Fig:ksksco}(b). The
fit quality is good ($\chi^2 /ndf = 86/97$). The peak around 1.3 GeV is suppressed due to
the destructive interference between $f_2(1270)$ and $a_2(1320)$ and the dip between 
$f_2(1270)/a_2(1320)$ and $f^{'}_2(1525)$ is well reproduced. The number of fitted $f_0(1710)$
events is $4058\pm 820$, which has $\approx 5$~s.d. significance.
Its mass is consistent with a $J^{PC}=0^{++}$ glueball candidate, but it cannot be a pure
glueball if it is the same state as in $\gamma\gamma\to K^0_S K^0_S$.
 
\begin{table}[!htb]
\begin{tabular}{|c|c|c|c|c|c|c|}
\hline 
Fit & \multicolumn{2}{c|}{No interference} & \multicolumn{2}{c|}{Interference} &  \multicolumn{2}{c|}{} \\
\cline{1-5}
\raisebox{0.ex}[0.6ex]{$\chi^2/ndf$} & \multicolumn{2}{c|}{\raisebox{0.ex}[0.6ex]{96/95}} & 
\multicolumn{2}{c|}{\raisebox{0.ex}[0.6ex]{86/97}} & 
\multicolumn{2}{c|}{\raisebox{1.7ex}[0.7ex]{PDG 2007 Values}} \\[-0.1ex]
\hline
in MeV & Mass & Width & Mass & Width & Mass & Width \\
\hline
$f_2(1270)$ & & & $1268\pm 10$ &  $176\pm 17$ &  $1275.4\pm 1.1$ &  $185.2^{+3.1}_{-2.5}$\\
\cline{1-1}\cline{4-7}
\raisebox{0.ex}{$a_2^0(1320)$} & \raisebox{1.7ex}[0.8ex]{$1304\pm 6$} & \raisebox{1.7ex}[0.8ex]{$61\pm 11$} & \raisebox{0.ex}{$1257\pm 9$}
 & \raisebox{0.ex}{$114\pm 14$} & \raisebox{0.ex}{$1318.3\pm 0.6$} & \raisebox{0.ex}{$107\pm 5$} 
\\[-0.ex]
\hline
$f_2^\prime(1525)$ & $1523\pm 3^{+2}_{-8}$ & $71\pm 5^{+17}_{-2}$ &  $1512\pm 3^{+2}_{-0.6}$ &  $83\pm
  9^{+5}_{-4}$ & $1525\pm 5$ &  $73^{+6}_{-5}$ \\
\hline
 $f_0(1710)$ & $1692\pm 6^{+9}_{-3}$ & $125\pm 12^{+19}_{-32}$ &  $1701\pm 5^{+5}_{-3}$ & $100\pm
 24^{+8}_{-19}$ & $1724\pm 7$ & $137\pm 8$  \\
 \hline
\end{tabular}
 \caption{\it Fitted masses and widths for $f_2(1270)$, $a_2^0(1320)$,
$f^{'}_2(1525)$ and $f_0(1710)$
                                                                  from the incoherent and coherent     
fits compared to PDG.       The first error is statistical.
For $f^{'}_2(1525), f_0(1710)$ the second errors are systematic uncertainties.
 }
\label{tab:limits}
\end{table}
 
 \begin{wrapfigure}{r}{0.5\columnwidth}
 \centerline{\includegraphics[width=0.45\textwidth]{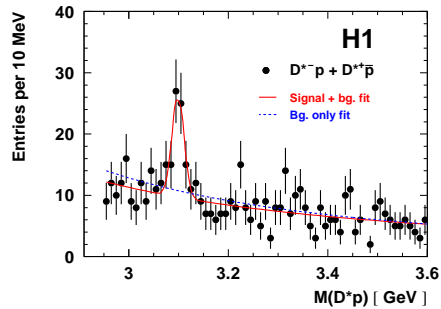}}
 \vspace{-0.7cm}  
\caption{ $M(D^{*\pm}p^{\mp})$ from H1 DIS HERA~I, compared with fit results where 
both signal and background components are included and where only background is included. }
                                               
                    \label{Fig:h1thetac}
                                                
\end{wrapfigure}

The          masses and widths obtained from both fits                                                   
are shown in Table \ref{tab:limits} and compared to           
PDG~\cite{PDG}.
                                     The no-interference
fit   yields a narrow width for the combined $f_2(1270)/a_2(1320)$ peak, as was
also seen by L3~\cite{L3}.
The fit with interference yields widths close to the PDG values for all observed resonances.
The $a_2^0(1320)$ mass is below the PDG value.           
                                    The $f^{'}_2(1525)$ and $f_0(1710)$
masses are somewhat below     PDG        with uncertainties comparable with the PDG ones. A fit
without $f_0(1710)$ is strongly disfavoured with $\chi^2 /ndf = 162/97$.
 
%




\section{Charm pentaquark search in the $D^* p$ system}

A narrow exotic baryon with strangeness +1 around 1530 MeV decaying into $K^+ n$ or 
$K^0 p$ was seen by various experiments and attributed to the $\Theta^+ =uudd\bar s$
pentaquark state predicted by Diakonov et al.\cite{diakonov}. 
If a   strange pentaquark            exists,                   charmed pentaquarks,
$\Theta^0_c =uudd\bar c$, could also exist.                                             
If $M(\Theta^0_c) > M(D^*)+M(p)=2948$~MeV, it can decay 
to $D^{*\pm}p^{\mp}$.

The H1 Collaboration saw~\cite{H1}                                                           
 in a DIS HERA~I
sample of $\approx 3400~D^{*\pm}\to D^0\pi^{\pm}_S\to K^{\mp}\pi^{\pm}\pi^{\pm}_S$                   
                                     a narrow signal of $50.6\pm 11.2$ events        
                                                     in the $D^{*\pm} p^{\mp}$ invariant mass at
$3.1$~GeV (Fig.\ref{Fig:h1thetac})                                                    
with     a   width     consistent with the mass resolution                                
     and a rate of  $\approx 1\%$ of the visible $D^*$ production.                      
 
ZEUS searched for a  
$\Theta^0_c$ signal in the $D^{*\pm} p^{\mp}$ mode with the full HERA~I PHP + DIS data sample~\cite{thetac}.
Clean $D^{*\pm}$ signals were seen in the $\Delta M~=~M(D^{*\pm})-~M(D^0)$ plots.
Two $D^{*\pm}\to~D^0\pi^{\pm}_S$      
          decay channels were used with
 $D^0\to~K^{\mp}\pi^{\pm}$ and               
 $D^0\to K^{\mp}\pi^{\pm}\pi^+\pi^-$.               
The $\Theta^0_c$ search was performed in the kinematic range                                         
                                        $|\eta (D^*)|< 1.6$ and                            
 $p_T (D^*) > 1.35 (2.8)$~GeV and with $\Delta M$ values between 0.144 - 0.147
(0.1445 - 0.1465) GeV for the  
$K\pi\pi$ ($K\pi\pi\pi\pi$) channel.                                                    
                         In these       bands                                                      
         $\approx 62000~D^*$'s were obtained                          
                                             after subtracting wrong-charge combinations
with       charge $\pm 2$ for the $D^0$ candidate and $\pm 1$ for the $D^*$ candidate.
Selecting DIS events with $Q^2 > 1$~GeV$^2$ yielded smaller, but cleaner $D^*$ signals with  
         $\approx 13500~D^*$'s.

 
 
 
 
 
 \vspace*{-1.5cm}      
 \begin{wrapfigure}{r}{0.5\columnwidth}
 \centerline{\includegraphics[width=0.45\textwidth]{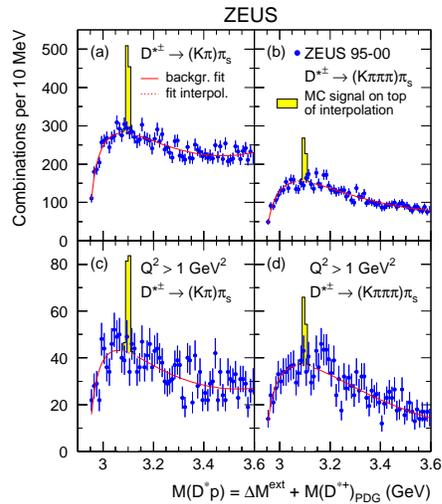}}
\caption{$M(D^{*\pm}p^{\mp})$ from ZEUS HERA~I. 
Solid curves are fits to a background function. Shaded historgams are MC $\Theta^0_c$
signals, normalised to $\Theta^0_c / D^* = 1\%$, on top of the background fit.  }
                                               
                    \label{Fig:zeusthetac}
                                                
\end{wrapfigure}

 \vspace*{+1.5cm}      
Protons were selected with momentum $P(p) > 0.15$~GeV. To reduce the pion and kaon background,
a   parameterisation of the expected $dE/dx$ as a function of $P/m$ was obtained using
tagged protons from $\Lambda$ decays and tagged pions from $K^0_S$ decays. The $\chi^2$
probability of the proton hypothesis was required to be above $0.15$. 
 
Figure \ref{Fig:zeusthetac}
                 shows the $M(D^{*\pm}p^{\mp})$
  distributions for the $D^0\to K\pi$ (left) and $D^0\to K\pi\pi\pi$ (right) channels for the full
 (up) and the DIS (down) samples.                                                         
No narrow
 signal is seen in any of the            distributions.                                       
$95\%$ C.L. upper limits on the fraction of $D^*$ mesons originating from  $\Theta^0_c$ decays,               
$R(\Theta^0_c\to D^* p/D^*)$,             were calculated
in a signal window $3.07 < M(D^* p) < 3.13$~GeV for the $K\pi\pi$ and $K\pi\pi\pi\pi$ channels.
                         The $M(D^* p)$ distributions were fitted                           to
 the form $x^a e^{-bx+cx^2}$, where $x=M(D^* p)-M(D^*)-m_p(PDG)$.                             
The number of reconstructed $\Theta^0_c$ baryons was estimated by subtracting the background
function  from the observed number of events in the signal window, yielding        
$R(\Theta^0_c\to D^* p/D^*) < 0.23\%$ and $ < 0.35\%$ for the full and DIS combined two channels.
A visible rate of $1\%$ for this fraction                                      
                                                                      is excluded by
 9 s.d. (5 s.d.) for the full (DIS) combined sample.
The acceptance-corrected rates are, respectively, $0.37\%$ and $0.51\%$.
The $95\%$ C.L. upper limit  on the fraction of charm quarks fragmenting to $\Theta^0_c$ times the
branching ratio    $\Theta^0_c\to D^* p$ for the combined two channels is
$f(c\to\Theta^0_c)\cdot B_{\Theta^0_c\to~D^* p} < 0.16\%$ ($ < 0.19\%$) for the full (DIS) sample.

\begin{wrapfigure}{r}{0.5\columnwidth}
\centerline{\includegraphics[width=0.45\textwidth]{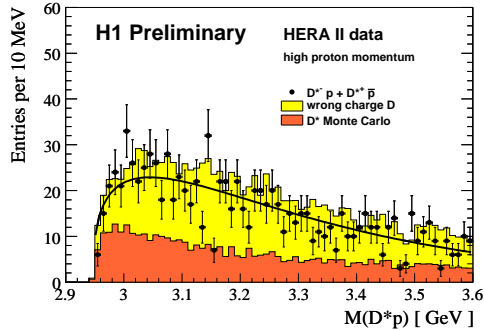}}
\caption{$M(D^{*\pm}p^{\mp})$ from H1 DIS HERA~II. The solid line is a
background parametrisation.}
                                               
                    \label{Fig:h1newthetac}
                                                
\end{wrapfigure}

In a HERA~II DIS data sample that is $\approx 4$ times larger than the HERA~I sample, H1 does
not see any significant peak at $3.1$~GeV (Fig.\ref{Fig:h1newthetac}).                           
                                           A preliminary $95\%$ C.L.             for the ratio
of $D^* p$ to $D^*$ is $0.1\%$.

 \vspace{-0.2cm}  
\section{Excited charm and charm-strange mesons}

 \vspace{-0.2cm}  
The large charm production               at HERA allows to search for excited charm states.
ZEUS studied the orbitally excited states $D_1 (2420)^0\to D^{*\pm}\pi^{\mp}$ ($J^P = 1^+$),
$D^*_2 (2460)^0\to D^{*\pm}\pi^{\mp} , D^{\pm}\pi^{\mp}$ ($J^P = 2^+$) and
$D_{s1}(2536)^{\pm}\to D^{*\pm} K^0_S , D^{*0}K^{\pm}$ ($J^P = 1^+$) and searched for the
radially excited state $D^{*'}(2640)^{\pm}\to D^{*\pm}\pi^+\pi^-$ ($J^P = 1^-$ ?) with
a   HERA~I PHP + DIS sample\cite{excited}.    ~
 
 \vspace{-0.6cm}  
 \hspace{+8.0cm}  
             
 \vspace{-0.8cm}  
 \begin{wrapfigure}{r}{0.5\columnwidth}
\centerline{\includegraphics[width=0.40\textwidth]{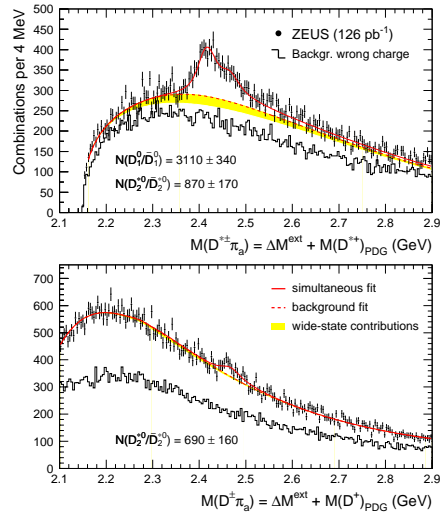}}
\caption{$M(D^{*\pm}\pi_a)$ and $M(D^{\pm}\pi_a)$ distributions. Solid curves are simultaneous
fit; dashed curves are background; histograms are wrong-charge combinations.}
                                               
                    \label{Fig:dss}
 
\end{wrapfigure}

 \vspace{+0.5cm}  
A large sample of events has been collected with the ground state charm mesons $D^{*\pm}, D^0, D^{\pm}$.
 The number of $D^{*\pm}$ mesons was obtained by subtracting the wrong charge
background. The number of $D^{\pm}\to K^{\mp}\pi^{\pm}\pi^{\pm}$         
  and $D^0(\bar D^0)\to K^{\mp}\pi^{\pm}$ was extracted from  fits 
to a modified Gauss function, $Gauss^{mod}\sim exp(-0.5x^{1+\frac{1}{(1+0.5x)}})$,          
where $x~=~(M-~M_D)~/\sigma$,                            plus a background
function. For the $D^*$, both $D^0$ decay modes to $K\pi$ and $K\pi\pi\pi$ were used.
 
\subsection{Excited charm mesons}

To reconstruct the excited charm mesons, a $D^{*\pm}$ or $D^{\pm}$ candidate was combined with a pion
of opposite charge, $\pi_a$.
Figure \ref{Fig:dss} shows the ``extended" mass difference distributions 
$M(D^{*\pm}\pi_a) - M(D^{*\pm}) + M(D^*)_{PDG}$ (upper plot) and 
$M(D^{\pm}\pi_a) - M(D^{\pm}) + M(D)_{PDG}$ (lower plot).             
          A clear excess is seen in $M(D^{*\pm}\pi^{\mp}_a)$ around the
$D^0_1 /D^{*0}_2$ mass region. A small excess near the $D^{*0}_2$ mass is seen in   
$M(D^{\pm}\pi^{\mp}_a)$.     No excess      is seen for wrong charge combinations, where
$D^* (D)$ and $\pi_a$ have the same charge.
 
To distinguish between the $D^0_1$ and $D^{*0}_2$, the helicity angular distribution,
parametrised as $dN/d\cos\alpha\approx 1 + h\cos^2\alpha$,
was used. Here $\alpha$ is the angle between the $\pi_a$ and
$\pi_S$ momenta in the $D^*$ rest frame. The helicity parameter $h$ is predicted~\cite{HQET} to be
$3 (-1)$ for pure D-wave $D^0_1$ ($D^{*0}_2$).
 
\begin{wrapfigure}{r}{0.5\columnwidth}
\centerline{\includegraphics[width=0.45\textwidth]{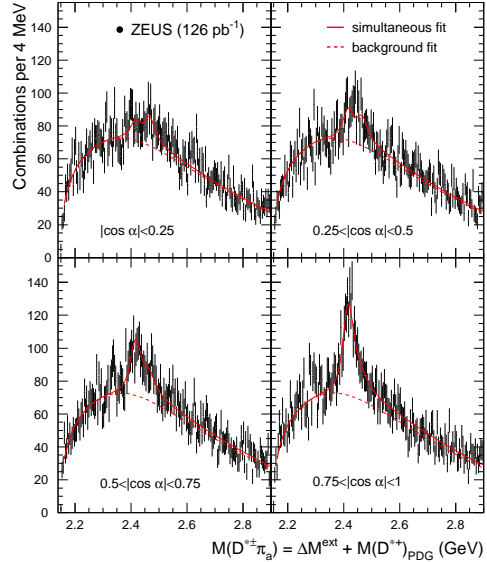}}
 \vspace{-0.5cm}  
 \hspace{+0.2cm}  
\caption{$M(D^{*\pm}\pi_a)$ distributions in 4 helicity intervals.      }
                                               
                    \label{Fig:dsshel}
                                                
\end{wrapfigure}

Figure \ref{Fig:dsshel} shows the $D^{*\pm}\pi_a$ ``extended" mass difference                 
                                      in 4 helicity $|\cos\alpha|$ intervals.
The $D^0_1$ contribution increases with $|\cos\alpha|$ and dominates for $|\cos\alpha| > 0.75$.
A simultaneous fit was performed to the 4 helicity regions of
Fig.\ref{Fig:dsshel} and to the $M(D\pi)$ distribution of Fig.\ref{Fig:dss}. The data is
described well with 15 free parameters (signal yields, masses, $D^0_1$ width and helicity).
The fitted masses agree with PDG. The fitted $D^0_1$ width is
             $53.2\pm 7.2(stat.)^{+3.3}_{-4.9}(syst.)$~MeV
compared to $20.4\pm 1.7$~MeV of PDG. The fitted $D^0_1$ helicity
 ($5.9^{+3.0}_{-1.7}(stat.)^{+2.4}_{-1.0}(syst.)$) is consistent with a pure D-wave.
 
 \newpage
 \vspace{-1.0cm}  
\subsection{Excited charm strange mesons}

\begin{wrapfigure}{r}{0.5\columnwidth}
\centerline{\includegraphics[width=0.40\textwidth]{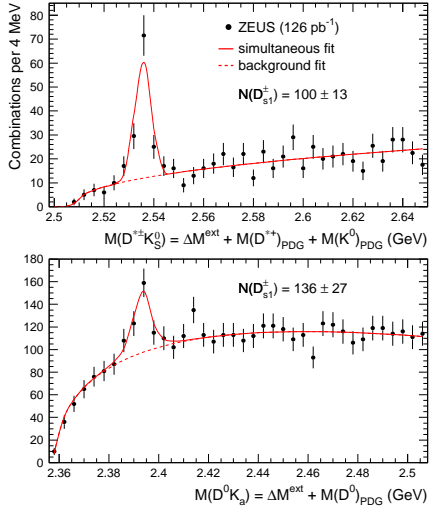}}
\caption{$M(D^{*\pm}K^0_S)$ and $M(D^0 K^{\pm})$ distributions. Solid curves are simultaneous
fit; dashed curves are background.}                                                        
                                               
                    \label{Fig:ds1}
                                                
\end{wrapfigure}

 \vspace{+0.5cm}  
To reconstruct the $D^{\pm}_{s1}\to D^{*\pm}K^0_S$ decays, a $D^{\pm}_{s1}$ candidate was
formed by combining a $D^*$ candidate with a reconstructed $K^0_S$ of the same event.
Figure \ref{Fig:ds1} (upper plot) shows the ``extended" mass difference distribution 
$M(D^{*\pm} K^0_S) - M(D^{*\pm}) + M(D^*)_{PDG} + M(K^0)_{PDG}$. 
 A clear $D_{s1}(2536)^{\pm}$ signal is seen.
The decay mode $D^{\pm}_{s1}\to D^{*0}K^{\pm}$ is reconstructed from          the "extended"
mass difference $M(D^0 K_a) - M(D^0) + M(D^0)_{PDG}$. A nice $D^{\pm}_{s1}$ signal is
seen (Figure \ref{Fig:ds1} lower plot) at a mass shifted down by $\approx 142$~MeV from the
$D^{\pm}_{s1}$ mass. The signal is a feed-down from $D^{\pm}_{s1}\to D^{*0}K^{\pm}$ with
$D^{*0}\to D^0\pi^0 , D^0\gamma$. 
An unbinned likelihood fit was performed using simultaneously values of $M(D^0 K_a)$, 
$M(D^{*\pm}K^0_S)$ and $\cos\alpha$ for the $D^{*\pm}K^0_S$ combinations. Yields and widths
of both signals and the $D^{\pm}_{s1}$ mass and helicity parameter were free parameters of
the fit. The fitted $D_{s1}$ helicity parameter is  
$h(D^{\pm}_{s1})=-0.74^{+0.23}_{-0.17}(stat.)^{+0.06}_{-0.05}(syst.)$. It is inconsistent with a pure 
$J^P = 1^+$ D-wave and is barely consistent with a pure $J^P = 1^+$ S-wave, indicating 
a significant $S-D$ mixing.

\begin{wrapfigure}{r}{0.5\columnwidth}
 \centerline{\includegraphics[width=0.50\textwidth]{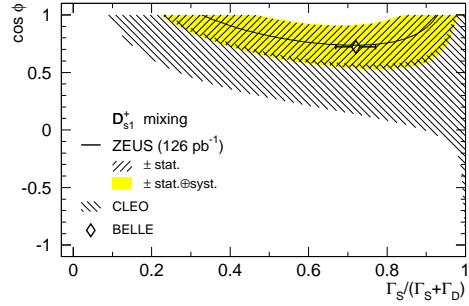}}
\caption{$\cos\phi$ vs. $\Gamma_S /(\Gamma_S + \Gamma_D)$ for $D_{s1}^+\to~D^{*+}K^0_S$ decay.}      
                                               
                    \label{Fig:Sfrac}
                                                
\end{wrapfigure}

The helicity angular distribution form of a   $1^+$ state for any D- and S-wave mixing is:
          $dN/d\cos\alpha\approx r + (1-r)(1+3\cos^2\alpha)/2 + \sqrt{2r(1-r)}\cos\phi
           (1-3\cos^2\alpha)$, where 
            $r=\Gamma_S/(\Gamma_S~+~\Gamma_D)$,              
        $\Gamma_{S/D}$  is the S/D wave partial width and
        $\phi$ is relative phase between the 2 amplitudes,
            $\cos\phi = \frac {(3-h)/(3+h) - r} {2\sqrt{2r(1-r)}}$.
 Figure \ref{Fig:Sfrac} shows a range, restricted by the measured $h(D^{\pm}_{s1})$ value
and its uncertainties, in a plot of $\cos\phi$ versus $r$. The measurement suggests a
significant contribution of both D- and S-wave amplitudes to the 
$D_{s1}(2536)^{\pm}\to D^{*\pm} K^0_S$ decay. The ZEUS range agrees nicely with the BELLE result
and roughly with the CLEO measurement.
 
 \vspace{+1.5cm}  

 \vspace{-1.5cm}  
 \hspace{+8.0cm}  

 \vspace{-0.5cm}  
             
 \newpage
\section{Branching ratios and fragmentation fractions}
 
Using the ZEUS measured fractions $f(c\to~D^{*+})$ and $f(c\to D^{+})$\cite{fraction},
the following decay rate ratios were derived:
$\frac{B_{D_{2}^{*0}\to D^{+}\pi^-}}
      {B_{D_{2}^{*0}\to D^{*+}\pi^-}}=
2.8\pm~0.8^{+0.5}_{-0.6}$ (PDG:        
 $2.3\pm0.6$);      
$\frac{B_{D_{s1}^+ \to D^{*0} K^+}}
{B_{D_{s1}^+ \to D^{*+} K^0}}=
2.3\pm~0.6\pm~0.3$
 (PDG: $1.27\pm0.21$).
 
Assuming isospin conservation for $D^0_1$ and $D^{*0}_2$ and 
$B_{D^+_{s1}\rightarrow D^{*+}K^0}+
B_{D^+_{s1}\rightarrow D^{*0}K^+}=1$
yields
a strangeness suppression of excited $D$ mesons 
$f(c\to D^{+}_{s1})/f(c\to D^0_1)=0.31\pm0.06 (stat.)^{+0.05}_{-0.04}(syst.)$.
 
In Table \ref{tab:fractions} the ZEUS fragmentation fractions of the excited charm mesons are
compared with   $e^+ e^-$ values. The results are consistent within errors.

DELPHI saw a narrow peak in $D^{*\pm}\pi^+\pi^-$ at $2637$~MeV~\cite{DELPHI} and attributed
it to a radially excited $D^{*'\pm}$. No signal was seen in ZEUS and a $95\%$ C.L. upper limit of
$f(c\to D^{*'\pm})\cdot B_{D^{*'\pm}\to D^{*+}\pi^+\pi^-} < 0.4\%$ was set, compared to the
weaker limit of OPAL ($0.9\%$)~\cite{OPAL}.

\begin{table}[!htb]
\begin{tabular}{|c|c|c|c|} \hline
& $f(c\to D^0_1) [\%]$ & $f(c\to D^{*0}_2) [\%]$ & $f(c\to D^+_{s1}) [\%]$ \\
\hline
\hline
     ZEUS  
&      $3.5\pm 0.4^{+0.4}_{-0.6}$ 
&      $3.8\pm 0.7^{+0.5}_{-0.6}$ 
&      $1.11\pm 0.16^{+0.08}_{-0.10}$  \\
\hline
OPAL
& $2.1\pm 0.8$
& $5.2\pm 2.6$
& $1.6\pm 0.4\pm 0.3$ \\
\hline
ALEPH
&
&
& $0.94\pm 0.22\pm 0.07$ \\
\hline
\end{tabular}
 \caption{\it The fractions of $c$ quarks hadronising into $D^0_1$, $D^{*0}_2$ and
$D^+ _{s1}$ mesons.}                
\label{tab:fractions}
\end{table}

\begin{footnotesize}

 
 
%
 
\end{footnotesize}
 
 
\end{document}